\documentclass[pra,aps,twocolumns,superscriptaddress,showpacs,footnotes]{revtex4}

\begin{document}

\title{Quantum non-locality - It ain't necessarily so...}

\author{Marek \.Zukowski}
\address{Institute of Theoretical Physics and Astrophysics, University of Gda\'nsk,
ul. Wita Stwosza 57, PL-80-952 Gda\'nsk, Poland}

\author{{\v C}aslav Brukner}
\address{Faculty of Physics, University of Vienna,
Boltzmanngasse 5, A-1090 Vienna, Austria;} \address{Institute for Quantum Optics and Quantum Information,
Austrian Academy of Sciences, Boltzmanngasse 3, A-1090 Vienna,
Austria}

\begin{abstract}

Bell's theorem is  50 years old. Still there is a controversy about its implications. Much of it has its roots in confusion regarding  the premises  from which the theorem can be derived. Some  claim that a derivation of Bell's inequalities requires just
locality assumption, and nothing more. Violations of the inequalities are then interpreted as  ``nonlocality'' or ``quantum nonlocality''. We show that such claims are unfounded and that every derivation of Bell's inequalities requires a premise -- in addition to locality and freedom of choice -- which is either assumed tacitly, or unconsciously, or is embedded in a single compound condition (like Bell's ``local causality''). The premise is equivalent to the assumption of existence of additional variables which do not appear in the quantum formalism (in form of determinism, or joint probability for outcomes of all conceivable measurements, or ``additional causes`'', or ``hidden variables'', ``complete description of the state'' or  counterfactual definiteness, etc.). A certain irony is that perhaps the main message of violation of Bell's inequalities is that our notion of locality should be based on an operationally well-defined no-signalling condition, rather than on local causality.

\end{abstract}


\maketitle

The terms `nonlocality' or `quantum non-locality' are buzzwords in foundations of quantum mechanics and quantum information. Most of scientists treat these terms as a more handy expression {\em equivalent} to the clumsy ``violation of Bell's inequalities''. Unfortunately, some treat them seriously. Even more unfortunately Bell himself used such terms in later works ~\cite{Bell1980,BELL-BEABLES} \footnote{In  Ref. \cite{BELL-BEABLES} a chapter is called `Locality inequality', and it refers to the Bell-CHSH inequality. Consistently with this, in Ref.~\cite{Bell1980,BELL-BEABLES} Bell writes that the famous inequalities imply non-locality. However, such statements are not in all works of Bell. Even in Ref.~\cite{Bell1980}, in the last paragraph, he writes the following, as one of the logically possible implications of violations of the inequalities: `The lesson of quantum mechanics is not to look behind the predictions of the formalism', and further `... it may be that Bohr's intuition was right -- in that there is no reality below some 'classical' 'macroscopic' level'.}. 

There are two approaches to Bell's theorem which we would like to critically analyze: 
\begin{itemize}
\item
(A) Derivation of a Bell inequality relies only on the premise of locality, and nothing more.
\item
(B) Derivation of Bell inequality does rely on the additional premise of determinism, or any of the notions listed in the abstract. However, these notions can be derived from the premises of locality, the freedom of an experimenter to choose the setting of his/her local apparatus and quantum predictions, and nothing more. 
\end{itemize}
Consequently, violations of Bell's inequalities  in either approach is interpreted as a demonstration of `non-locality''.
(As it is widely accepted that it is difficult to dismiss the `freedom of choice' assumption this aspect of Bell's theorem will be not discussed.) Supporters of (A), state that it is not clear what realism,  in the context of Bell's theorem, should mean, and that those who use this notion most probably confuse it with determinism -- the notion that outcomes for all possible measurements are pre-determined. 
Additionally, they claim that determinism, or the other versions of the assumption listed in the abstract, needs not to be assumed neither in stating nor in proving Bell's inequalities, and thus cannot be the issue. Followers of (B) see this differently. They use pre-determined values in derivation of Bell's inequalities but claim that it is an implication of locality, freedom of choice, and experimentally observed perfect correlations. Hence, they basically follow the Einstein-Podolsky-Rosen (EPR) argument~\cite{EPR}.

We show that Bell's condition of ``local causality''~\cite{Bell2004} which proponents of view (A) often adopt is {\em equivalent} to the assumption of existence of a (positive and normalized) joint probability distribution for the values of all possible measurements that could be performed on an individual system, no matter whether any measurement -- and which measurement -- is actually performed. This shows that local causality is a {\em compound} condition, which assumes the existence of causes which are not present in quantum mechanics, and that the causes act locally (i.e. their influence is bound to the future light cone). Hence, {\em locality is not the sole assumption} of view (A) based on local causality. This refutes the view (A). 

We refute the view (B): any attempts to derive determinism, or hidden-variables, via an EPR argument~\cite{EPR} are futile. The argument of EPR was based on an assumption, which is often not noticed or forgotten: {\em counterfactual definiteness}~\cite{MUYNK,MUYNK-2}. This assumption allows one to assume the definiteness of the results of measurements, which were actually {\em not} performed on a given individual system. They are treated as unknown, but in principle {\em defined} values.  {{} This is in a striking disagreement with quantum mechanics, and complementarity principle. Therefore, it cannot be glossed over as something not worth mentioning as an additional assumption.}

Our conclusion is that, despite suggestions to the contrary, one always assumes an additional premise, except locality and freedom, to derive any Bell inequality. The additional premise is either assumed tacitly, or unconsciously\footnote{ `Par ma foi! il y a plus de quarante ans que je dis de la prose sans que j'en susse rien ...' (Moliere, Le Bourgeois Gentilhomme, 1670)}, or is embedded in a single compound condition (like Bell's `local causality'), from which the inequalities are derived.

\section{Locality and nothing more?}

Bell's discussion of the premises behind his theorem, which differs substantially with respect to his original approach is given most extensively in his final published paper `La nouvelle cuisine'~\cite{Bell2004}. We will base our discussion on it, as the work is often referred to by advocates of (A). In this work Bell fully defines his new notion of local causality.

Consider two space-like separated parties, Alice and Bob. Either of them has a choice between a number of measurement settings. Denote, respectively, by $x$ and $y$ Alice's and Bob's choice of measurement settings and by $A$ and $B$ their
outcomes. Correlations of the outcomes are described by a conditional probability distribution $p(A,B|x,y)$. 

It is argued, that the situation described by  $p(A,B|x,y)$ may arise out of a statistical mixture of different situations traditionally labeled by $\lambda$ and sometimes called `causes'.
The probabilities therefore acquire the following form $p(A,B|x,y)=\int d\lambda \rho(\lambda) p(A,B|x,y,\lambda)$, where $\rho(\lambda)$ is a probability distribution. The properties of conditional probabilities allow one to put $ p(A,B|x,y,\lambda)= p(A|B,x,y,\lambda) p(B|x,y,\lambda)$. Next, local causality stated by Bell as `The direct causes (and effects) of [the] event are near by, and even the indirect causes (and effects) are no further away than permitted by the velocity of light', allows one to state that `what happens on Alice's side does not depend on what happens on Bob's side' and {\em vice versa}~\cite{Gisin}. This results in
$p(A|B, x, y, \lambda) = p(A|x,\lambda)$ and $p(B|x,y,\lambda)=p(B|y,\lambda)$. Finally, we obtain the general mathematical structure underlying derivations of all Bell's inequalities:
\begin{equation}
p(A,B|x,y)=\int d\lambda \rho(\lambda) p(A|x,\lambda) p(B|y,\lambda).
\label{condition}
\end{equation}

Proponents of (A) then argue that besides `locality' nothing more has been assumed in the derivation of equation ~(\ref{condition}). Obviously determinism is not required. The  condition~(\ref{condition}) holds also for any local stochastic (indeterministic, hidden variable) theory. While a deterministic theory specifies which outcome, exactly, will happen under a given  $\lambda$, a stochastic theory specifies only the probabilities for various outcomes that might be realized. (Note that the former is a special case of the latter, with the probabilities always being exclusively 1 or 0.). So, where, if at all, ``realism'' of any sort could be hidden in condition~(\ref{condition})?

Note first, that in the case of the {\em mixed} separable states, one may think that $\lambda$ specifies the ``actual'' quantum state in the probabilistic mixture. All this would agree with the formula~(\ref{condition}). However when the joint quantum state is an entangled one, {{} and especially if it is additionally pure}, and no additional $\lambda$'s are introduced, one cannot have a factorization like~(\ref{condition}), see e.g. Ref.~\cite{Gisin}, or appendix B. Thus the formula (\ref{condition}) is equivalent to the introduction of additional hidden parameters, $\lambda$'s, which are {\em not} present in  quantum theory. The $\lambda$, which enter Eq.~(\ref{condition}), can pop up under many guises such as,  e.g. `the physical state of the systems as described by any possible future theory'~\cite{Gisin}, `local beables', `the real state of affairs', `complete description of the state', etc.
Since $\lambda$ do not appear in quantum mechanics, thus they are (good old)  {\em hidden variables}. Anything on which one conditions probabilities, which gives {{}  {\em different structure} to formulas for probabilities than quantum mechanical formalism\footnote{{{} The basic quantum structure is $Tr\rho \Pi$}, where $\rho$ is the state and $\Pi$ is a projector.}}, is a hidden variable {\em per se}. Bell himself writes `$\lambda $ denote any number of hypothetical additional  variables needed to complete quantum mechanics in the way envisaged by EPR'~(\cite{Bell2004}, page 242). This sentence of Bell's is often forgotten by supporters of (A). As we will see next, we can try to introduce  $\lambda$'s not existent in quantum theory, but this will result in a condition which is stronger than locality alone.

The condition~(\ref{condition}) allows assignment of (positive and properly normalized) {\em joint} probabilities for the (local) values of the entire set of pairs of (local) measurements. Denote as $A_{x}$ the value of the outcome $A$ pertaining to the situation in which Alice chooses setting $x$, which, for example, in the Clauser-Horne-Shimony-Holt (CHSH)~\cite{CHSH} scenario has two possible values, denoted here as $1,2$, and similarly define $B_{y}$ ($y=1,2$). We can always introduce a joint probability 
\begin{eqnarray}
&p(A_{1}, A_{2},B_{1}, B_{2}|\lambda)&\nonumber\\ & = p(A|x=1,\lambda) p(A|x=2,\lambda)p(B|y=1,\lambda) p(B|y=2,\lambda)&.\\  \nonumber
\end{eqnarray} 
When the probabilities are 0 or 1 the model is deterministic, otherwise it is stochastic. Hence, we have defined a joint probability for all possible outcomes under all possible pairs of settings in the CHSH scenario. These outcomes include, for a single run of a Bell experiment, the actually measured ones and on an equal footing the ones which could have been potentially measured. Note that $\lambda$ is no more necessary here, as to get predictions for such a model it is enough to know $p(A_{1}, A_{2},B_{1}, B_{2})$, which is given by $\int d\lambda p(A_{1}, A_{2},B_{1}, B_{2}|\lambda)\rho(\lambda)$.  Starting from equation ~(\ref{condition}) one can introduce the joint probability $p(A_{1}, A_{2}, ..., B_{1}, B_{2},...)$ for an arbitrary number of settings. Note that the existence of such joint probability implies the existence of $p(A_{1}, A_{2}, ...)\geq 0,$  which has nothing to do with locality and is already in conflict with the Kochen-Specker theorem~\cite{KOCHEN}. 

Conversely, starting from the existence of the joint probability one can derive ~(\ref{condition}) for any pair of local settings. This is because in Kolmogorovian probability theory, which is the standard axiomatization of classical probability, if $\Omega$ is a probability space with probability measure $\rho(\lambda)$ and $A$ is a measurable set contained in $\Omega$, then indicator function $\chi_A(\lambda)$, which is $1$ if element (elementary event) $\lambda$ belongs to $A$, and $0$ when $\lambda$ belongs to the complement of $A$, that is  $\Omega \setminus A$, gives by the formula
$P(A)=\int_{\Omega}\chi_A(\lambda)\rho(\lambda)d\lambda$ the probability of the event $A$. The probability  
 $p(A_{1},A_{2},B_{1}, B_{2})$ can be modelled by $p(A_{1},A_{2},B_{1}, B_{2}|\lambda) = \chi_{A_1}(\lambda) \chi_{A_2}(\lambda) \chi_{B_1}(\lambda) \chi_{B_2}(\lambda) $, that is one can put  $p(A_{1},A_{2},B_{1}, B_{2}) = \int_{\Omega} \chi_{A_1}(\lambda) \chi_{A_2}(\lambda) \chi_{B_1}(\lambda) \chi_{B_2}(\lambda) \rho(\lambda) d\lambda$. Then by calculating the marginals, i.e. summing up the probabilities over the outcomes of the observables which are {\em not} measured in an actual experiment, one obtains: 
 $P(A,B|x,y)=\int_{\Omega}\chi_{A_x}(\lambda)\chi_{B_y}(\lambda)\rho(\lambda)d\lambda, $ where $\chi_{A_x}(\lambda)$ and $\chi_{B_y}(\lambda)$ are the indicator functions for the events $A_x$ and $B_y$ respectively. This is of course, equivalent to equation ~(\ref{condition}). 

Thus, we see that local causality condition (\ref{condition}) is mathematically equivalent to the assumption of joint  probabilities, $p(A_1,A_2,B_1,B_2)$. The latter {\it is} a form of realism: complementary observables are treated as mere numbers (`c-numbers' in Dirac's terminology~\cite{DIRAC}). One can speak about the joint probability  of $A_1,A_2$ and $B_1,B_2$ only if in a theoretical construction these all  values coexist together independently of which experiment is actually performed on either side, and in this sense are ``real''. Expressing it differently, the existence of a proper distribution $p(A_1,A_2,B_1,B_2)$ means that we model everything with Kolmogorovian probabilities, which always have a lack of knowledge interpretation,  with  the underlying $\Omega$ is treated as a sample space. Indeed, according to the previous results due to Fine~\cite{Fine}, Hall~\cite{Hall}, Gill {\em et al.}~\cite{Gill} and others, the probabilities of a local stochastic model {\em can} always be understood as stemming from our ignorance about supposed local deterministic values. However, we do not need to insist on the existence of  deterministic values of $A_1,A_2$ and $B_1,B_2$, The supposed existence of their joint probabilities is sufficient for our argument. In fact, the entire discussion on local deterministic vs. stochastic models -- on which the proponents of view (A) build up their argument -- is irrelevant for the current discussion. This is most evident if one takes into account the Greenberger-Horne-Zeilinger (GHZ)~\cite{GHZ} type of argument, which involves only perfect correlations between {\em three} (or more) systems and thus no stochastic model can ever recover them. (Note, that it is symptomatic, that followers  of view (A) always base their argument on correlations between two systems.) Furthermore, assuming a fundamental local indeterminism (that is some fundamental finite irreducible stochasticity, which does not allow one to use deterministic models -- even as only a mathematical tool) would lead to different bounds of the Bell inequalities. We show this in Appendix A. Somewhat ironically, not {\em any}  assumption of realism but its most stringent version, namely determinism itself, must be invoked to obtain the proper bounds of Bell's inequalities.

{{} The following observation can be made.
Some researchers are willing to accept that an outcome measured on a single (local) system may be intrinsically probabilistic. However, they do not accept the same for correlations between outcomes measured on several such systems; the probabilities for correlations are always taken to be reducible to probabilities for local outcomes in the form like (\ref{condition}). Confronted with
the experimental violation of Bell's inequalities,  they accept an inherently probabilistic explanation for an individual quantum system but hold fast to a pseudo-causal \footnote{That is, not obeying relativistic causality.} nonlocal one for correlations (e.g. by introducing faster-then-light influences between distant quantum systems). } 

\section{Derived realism?}

The school of thought (B) claims that one does need any version of the assumption of realism, or hidden variables to derive Bell's inequalities. However, it is purported, that everything  is deductible via EPR type reasoning. The current classic expositions of such line of thought can be found in Refs. \cite{NORSEN-1,NORSEN-2,TUMULKA}. For example, the argumentation in Ref.~\cite{TUMULKA} tries to oppose the results of the ``Free Will Theorem" of Ref.~\cite{CONWAY}.
To this end,  the following two logical statements are both claimed to be  true.
The first one is, that the logic of Bell's theorem is in the validity of the following implication  (we use the following terminology for the  structure of a logical implication: $assumption\Rightarrow thesis$):
\begin{equation}
freedom\hspace{1mm}  \& \hspace{1mm} QF\hspace{1mm}  \& \hspace{1mm}  locality \Rightarrow contradiction, \label{T-BELL}
\end{equation}
where $freedom$ stands for the freedom of choice of local measurement settings assumption, $locality$ for locality assumption and $QF$ for quantum formalism. On the other hand, the logic of the EPR paper supposedly allows to establish a yet another implication as valid:
\begin{equation}
freedom\hspace{1mm}  \&\hspace{1mm}  QF \hspace{1mm} \& \hspace{1mm} locality \Rightarrow determinism. \label{T-EPR}
\end{equation}
The validity of these statements is used to challenge the implication of Ref.~\cite{CONWAY}, which as put in Ref.~\cite{TUMULKA}, is 
\begin{equation}
freedom\hspace{1mm} \&\hspace{1mm} QF\hspace{1mm} \&\hspace{1mm} locality \hspace{1mm} \&\hspace{1mm} determinism\Rightarrow contradiction. \label{WILL}
\end{equation}

Obviously, if the implication (\ref{T-BELL}) is valid, {{} the assumption which leads to it cannot hold}. This is  because {\em contradiction} means a false statement (like $2\geq 2\sqrt{2}$). The rules of Aristotelian two-valued logic say that if the thesis of a {\em valid} (that is, true)  implication is false, then its assumption must be false too ({\em modus tollens}, see the truth table of logical implication). Thus the assumption of (\ref{T-EPR}), as it is the same as of (\ref{T-BELL}), must be false and one cannot `determine' whether determinism is true, from the validity of the implication   (\ref{T-EPR}), as from false statements one can derive via  a valid implication  both true and false statements (see again the truth table).  

Of course the first implication~(\ref{T-BELL}) is highly appealing to proponents of (A), while the second one~(\ref{T-EPR}) for supporters of (B). Alone the first implication would mean {\em non-locality} of quantum mechanics (provided freedom holds). But,  {\em  ``it ain't necessarily so"}~\cite{GERSHWIN}.
It is the implication (\ref{WILL}), which
correctly describes the situation, see below.

\subsection{EPR Reasoning}

The works \cite{NORSEN-1} and \cite{NORSEN-2} are based on an assumption of the validity of the EPR reasoning. For example in Ref.~\cite{NORSEN-2} one can find that `the existence
of these local, deterministic, non-contextual hidden variables [...]  is not simply assumed, but is inferred  from Locality plus a certain subset of the quantum mechanical predictions, using (in
essence) the EPR argument.' The work \cite{NORSEN-1} is aimed at showing that:  `The
hidden variables posited by Bell are not an `ad hoc assumption' but, rather, a logical implication of locality.'

The reasoning of EPR is often presented in the form of (\ref{T-EPR}), but this is wrong.
The logical structure  of the EPR argumentation about elements of reality is
\begin{eqnarray}
freedom \hspace{1mm} \&\hspace{1mm} QF\hspace{1mm} \&\hspace{1mm} locality \hspace{1mm} \& \hspace{1mm} counterfactual  \hspace{1mm} definitness &&\nonumber \\
\Rightarrow for\hspace{1mm} specific\hspace{1mm}observables\hspace{1mm} elements\hspace{1mm}of\hspace{1mm}reality\hspace{1mm}exist \hspace{1mm}for\hspace{1mm}the\hspace{1mm}
EPR\hspace{1mm}state. &&\label{EPR-COR}\nonumber \\
\end{eqnarray}
The thesis of the implication is true, provided one does not make an unfounded generalization of it  to {\em arbitrary} observables and {\em arbitrary} states. Indeed, for the EPR state and momentum and position observables ($P$ and $Q$), elements of reality seem to be a consistent notion. However this is arrived at by considering two situations of which only one can be the case in the given run of the experiment (measuring either $P$ or $Q$, page 780 of \cite{EPR}). This is counterfactual definitness at work.

 As a matter of fact such a generalization mentioned above  is the effective claim of EPR, as they aimed to prove incompleteness of the entire theory of quantum mechanics. A missing (hidden) part of the theory would be according to them {\em inherently deterministic ``elements of reality''}. By EPR, since they correspond to values of observables which can be predicted with certainty (i.e., with probability equal to unity), an act of measurement just displays the previously hidden values. 
However, there is a  logical flaw in the paper of EPR.  General existence of elements of reality is just their conjecture, based on just {\em one} example, while they claim that this is their thesis, holding always. Why conjecture? EPR  introduced elements of reality just for  the `original EPR' state.  And this was successful, that is it  did not lead to a contradiction, only because they limited themselves to specific observables, which do not exhaust all possible ones. Note here, that  for  different observables and the original EPR state,  in Ref.~\cite{BANASZEK}, one can find a proof of internal inconsistency of the EPR concepts.

In their conclusions EPR tacitly assumed that one can establish local elements of reality for all states with perfect correlations, and for all observables. This is wrong, as it can be directly shown in the case of the GHZ states. The GHZ  reasoning shows that the very definition of elements of reality must be wrong, as it implies contradictory values  (a $1=-1$ contradiction in the case of GHZ correlations, or $2\sqrt{2}\leq 2$ in the case of the CHSH version of the Bell theorem). This means that the thought-provoking paper of EPR does not contain a valid general statement on quantum mechanics! One could only {\em counterfactually} wonder what would have been the views of EPR, had the GHZ paper appeared before 1935.

Note further that the {\em thesis}  of Bohm's version of the EPR implication \cite{BOHM}
\begin{eqnarray}
&freedom \hspace{1mm} \&\hspace{1mm} QF\hspace{1mm} \&\hspace{1mm} locality \hspace{1mm} \& \hspace{1mm} counterfactual  \hspace{1mm} definitness &\nonumber \\
&\Rightarrow  elements\hspace{1mm}of\hspace{1mm}reality\hspace{1mm}exist \hspace{1mm}for\hspace{1mm}the\hspace{1mm}
two-spins-1/2-singlet\hspace{1mm}state, &\label{EPR-COR-2}\nonumber\\
\end{eqnarray}
is plainly {\em{not}} true. {{}  The  thesis of this implication leads} to a $2\geq2\sqrt{2}$ contradiction,  if one forms  the CHSH inequality for the elements of reality, and uses again quantum predictions. {{} Of course, counterfactual definitness was a tacit assumption in Bohm's reasoning.}
  
Comparing the relation of (\ref{EPR-COR}) with the views of proponents of (B), one sees the following differences with the simplified description of  the EPR result given by (\ref{T-EPR}). The assumption of {\em freedom} is sometimes thought to encompass also {\em counterfactual definitness}, and pre-determined values are thought to be derivable for all states with perfect correlations. The second is invalidated by the Bell theorem and even more strikingly by the GHZ reasoning. However, the basic problem here is that freedom does not encompass counterfactual definitness. This is evident, for example, in the case of the CHSH inequality where we discuss four situations (four possible pairs of local settings), only {\em one} of which can actually occur for the given pair or particles (run of the experiment). However, the proper bounds of inequalities (see appendix A) are derived by an algebraic manipulation of values for {\em a single pair}, for the actual situation and three counterfactual ones (``had one or both the observers chosen different settings'')\footnote{ For a more extended analysis see, e.g., the recent Ref. \cite{STAIRS}.}. The results that would have been obtained in such cases are then treated as unknown, but nevertheless definite real numbers ($\pm 1$ in the case of the CHSH scenario).{{} This is counterfactual definitness {\em per se}.} In this way an effective determinism enters.

Counterfactual definiteness is directly irreconcilable with Bohr's complementarity principle. The principle says that,  once a value of an observable is measured, one is not allowed to even speak about values for complementary observables. This is reflected by  quantum formalism, according to which probability distributions for complementary (non-commuting) observables are {{} in general} not defined, or rather they a {\em undefinable} within the theory. Thus, introduction of hidden variables, determinism, counterfactual definiteness etc. is not a minor point, a soft option, or something so obvious that may be treated as a tacit, indisputable assumption.
It goes directly against the very essence of quantum mechanics.

\section{Conclusions}

The terms `nonlocality' or `quantum nonlocality' suggest that there is some `spooky' faster-than-light influence between distant quantum systems (for a critical analysis see Ref.~\cite{Hall}). While the possibility of such an influence cannot be excluded, it is just a one-sided view. A failure of realism in all of its forms, for example, futility of considering joint probabilities for outcomes for the entire set of conceivable measurements or rejection of counterfactual  reasoning, is an equally valid option, just as a failure of both realism {\em and} locality.

In ``La Nouvelle Cuisine" Bell describes failure of {\em local causality} in the quantum world. This failure does not mean that we have to accept non-local causality. Individual events may have {\em spontaneous, acausal} nature. There seem to be no need to go beyond quantum mechanics within this aspect~\footnote{In his masterpiece book~\cite{PERES}, Peres, when discussing Bell's theorem, uses a phrase `non-locality is inescapable'. However, a careful reader would notice that he firmly assumes that counterfactual reasoning is acceptable (statements like had we measured a different observable than the actual one which was measured, we would have obtained a value, say $x$). Only using  a counterfactual reasoning measured and unmeasured, but potentially measurable, variables can be put on an equal footing. If one does not consider this assumption as disputable, non-locality is an inescapable consequence of Bell's theorem. In his later work with Fuchs~\cite{PERES-FUCHS}, he rejects non-locality. Also his well known slogan `unperformed experiments have no results' \cite{UNPERFORM} is a clear rejection of counterfactual definitness.}. Paradoxically, it was Einstein who reluctantly introduced the notion of spontaneous events, which might be after all the root of Bell's theorem. The lesson for future could however be that we should build the notion of locality on the operationally clear ``no-signalling'' condition -- the impossibility to transfer information faster-than-light. After all this is all what theory of relativity requires. 

{{} The moral of the story is that Bell's theorem, in all its forms, tells us {\em not} what quantum mechanics {\em is}, but what quantum mechanics {\em is not}.}

\section{Appendix A}

We show that inherently stochastic local hidden variable theories (that is, the theories with some fundamental finite irreducible stochasticity) lead to lower  bounds than the standard ones of Bell inequalities.
To this end we derive a Bell inequality for a local {\em strictly} stochastic theory. The latter is equivalent to the requirement that  all $p(A|x,\lambda)$ and $p(B|y,\lambda)$ are strictly different from 0 or 1. Therefore, suppose that there is inherent stochasticity parameter, $s>0$, which gives the following upper bound:
for any $\lambda$
\begin{equation}
s<p_s(X|z,\lambda)<1-s,
\label{BOUND}
\end{equation}
where $X=A,B,$ and $z=x,y$. As Bell's inequalities are in form of multilinear  combinations (functions) of the underlying probabilities, their bounds (maxima and minima) are at the border of the region of validity of the probabilities. In the standard case the borders are $0\leq p(X|z,\lambda)\leq1$. Thus,  the standard bounds cannot be reached in the case of probabilities satisfying (\ref{BOUND}). For example, take the CHSH inequality~\cite{CHSH}. With (\ref{BOUND}),
for every $\lambda$ one can introduce the expectation value $I(z, \lambda)=\sum_{x=\pm1}X p(X|z,\lambda)$ (we  assume here the usual spectrum of the Bell observables: $X=\pm1$).   
Obviously $-1+2s\leq I(z, \lambda)\leq 1-2s$. Thus, `renormalized' variables $I'=\frac{1}{1-2s}I$ give the usual of the CHSH inequality, that is 2.
However, we have
\begin{eqnarray}
&\langle 
I(x,\lambda)I(y,\lambda)+I(x,\lambda)I(y',\lambda)&\nonumber\\
&+I(x',\lambda)I(y,\lambda)-I(x',\lambda)I(y',\lambda)\rangle\leq 2(1-2s)^2<2.&\nonumber
\end{eqnarray}
The bound\footnote{ One can choose infinitesimally small but nonzero $s$ such that the new bound is arbitrarily close to 2 and still we have, in the strict mathematical sense, a non-deterministic theory. However, the different viewpoints (here determinism vs. nondeterminism with an infinitesimal $s$) would not be operationally distinguishable.} of the CHSH inequality in a local (intrinsically) stochastic theory is strictly smaller than 2. One must assume {\em both} locality and determinism in order to obtain the proper bounds for Bell's inequalities\footnote{Note that in order to obtain the usual bound it is enough to assume that e.g. results of Bob are {\em deterministic}, and  one of the results on Alice's side is deterministic too. Of course such an asymmetry of indeterminism is ridiculous. This is why  we assumed an irreducible stochasticity, $s$, at both sides of the experiment, for all settings.}.

\section{Appendix B: Alternative analysis}
{{} One can formulate the analysis of the assumption of local causality also in another alternative way.
The principle of local causality, $p(A|B,x,y,\lambda)=p(A|x, \lambda)$  and  $p(B|A,x,y,\lambda)=p(B|x, \lambda)$, implies in {\em  the case of correlated systems}:
\begin{itemize}
\item
An easily derivable relation
\begin{equation}
 p(A|x,y,\lambda)=p(A|x, \lambda),
\end{equation}
which a form of the {\em no-signaling condition}. {This is because under local causality  $p(A|x,y,\lambda)=\sum_B P(A,B|x,y,\lambda)=p(A|x, \lambda)\sum_B P(B|y,\lambda).$}
\item
Existence {\em at least two different}  values of `causes' $\lambda$, which makes $\lambda$  a non-trivial parameter (variable)  outside quantum formalism (i.e, a non-trivial hidden variable). 
\end{itemize}

Why the second implication? If one  Assumes just {\em one} common cause and local causality, this immediately leads to {\em no} correlations. To see this consider the following relations:  
\begin{equation}
P(A,B|x, y, \psi)= P(A|B,x, y, \psi) P(B|y, \psi)= P(A| x, \psi) P(B|y, \psi), 
\end{equation}
where is $\psi$ is the sole cause, which can be thought of  a pure (entangled) quantum state describing the preparation. Here we used first  probability rules and later local causality. The factorization means: {\em no} correlations.
 In order to describe correlations one {\em must} have at least {{} two} values for $\lambda$ and this implies that $\lambda$ is outside of the quantum formalism, as the only common cause, in the case of pure quantum states, allowed by quantum mechanics is the quantum state. Just one $\lambda=\psi$ cannot give us any correlations whatsoever.

Thus local causal theories (of correlated systems) are a subset of  non-trivial hidden variable theories (ones allowing for $\lambda$, outside of the quantum formalism), and local theories (which additionally impose $p(A|x,y, \lambda)=p(A|x,\lambda )$). That is, a local causal theory is a specific example  of a local hidden variable theory.

 Quantum mechanics  (QM) and quantum field theory (QFT) are no-signalling. In  QFT it  is assumed the for two gauge invariant observables, $O_1, O_2$ at two spatially separated space time points $\xi_1, \xi_2$
one has $[O_1(\xi_1),O_2(\xi_2)]=0$, and in QM one has for the results of a local observable $x$ defined on one of the local subsystems: $P(A|x,y,\psi)=P(A|x,\psi)$ (the meaning of the symbols is as above). Note that both QFT and QM are {\em not} local causal, but this is precisely the point of Bell's theorem. Briefly, if one wants to work outside local hidden variable theories, one cannot even formulate the local causality principle. Local causal hidden variable theories assume first of all non-trivial  hidden variables $\lambda$, and next that a form of locality principle is satisfied, in the form of $P(A|B, x,y, \lambda)=P(A|x, \lambda)$, etc. }

\section{Acknowledgements}
MZ is supported by a CHIST-ERA NCBiR project QUASAR. {\v C}B has been supported by the European Commission Project RAQUEL, the John Templeton Foundation, FQXi, and the Austrian Science Fund (FWF) through CoQuS, SFB FoQuS, and the Individual Project 2462.

\end{document}